%% file: conf00.tex
\documentclass[11pt]{article}
\usepackage{epsfig}
\usepackage{psfig}
\usepackage{graphicx}
\usepackage{rotating}
\usepackage{amsmath}
\usepackage{amssymb}
\usepackage{psfrag}
\usepackage{multirow}
\usepackage{pst-all}      
\usepackage{pst-poly}     
\usepackage{pst-coil}     
\usepackage{multido}      
\usepackage{pst-grad}
\newcommand{\BABARPubYear}    {04}

\newcommand{\BABARConfNumber} {037}
\newcommand{\SLACPubNumber} {10583}

\input pubboard/babarsym
 
\def\Bztorhozrhoz {\ensuremath{\Bz \to \rho^0 \rho^0 }\xspace}
\def\Btozz {\ensuremath{\Bz \to \rho^0 \rho^0 }\xspace}
\def\rhozrhoz {\ensuremath{\rho^0\rho^0 }\xspace}
\def\Bztorhoprhom {\ensuremath{\Bz \to \rho^+ \rho^- }\xspace}
\def\Bptorhozrrhop {\ensuremath{\Bp \to \rho^+ \rho^0 }\xspace}

\def\rhoz {\ensuremath{\rho^0}\xspace}
\def\rhop {\ensuremath{\rho^+}\xspace}
\def\rhom {\ensuremath{\rho^-}\xspace}

\long\def\inst#1{\par\nobreak\kern 4pt\nobreak
    {\it #1}\par\vskip 10pt plus 3pt minus 3pt}

\begin{document}
{\pagestyle{empty}


\begin{flushright}
\babar-CONF-\BABARPubYear/\BABARConfNumber \\
SLAC-PUB-\SLACPubNumber \\
August 2004 \\
\end{flushright}

\par\vskip 5cm

\begin{center}
\Large \bf Search for the Decay \boldmath\Btozz
\end{center}
\bigskip

\begin{center}
\large The \babar\ Collaboration\\
\mbox{ }\\
August 13, 2004
\end{center}
\bigskip \bigskip

\begin{center}
\large \bf Abstract
\end{center}
The \Btozz decay mode is searched for in a data sample of about 
227 million \upsbb decays collected with the \babar\ detector at the
\pep2\ asymmetric B factory at SLAC.  No significant signal is observed, 
and an
upper limit of $1.1\times  10^{-6}$ (90\% C.L.) on
the branching fraction is set.
Implications on the penguin contribution and 
constraints on the CKM angle $\alpha$ with $B\to\rho\rho$ 
decays are discussed. All results are preliminary.
\vfill
\begin{center}

Submitted to the 32$^{\rm nd}$ International Conference on High-Energy Physics, ICHEP 04,\\
16 August---22 August 2004, Beijing, China

\end{center}

\vspace{1.0cm}
\begin{center}
{\em Stanford Linear Accelerator Center, Stanford University, 
Stanford, CA 94309} \\ \vspace{0.1cm}\hrule\vspace{0.1cm}
Work supported in part by Department of Energy contract DE-AC03-76SF00515.
\end{center}

\newpage
} 

\input pubboard/authors_sum2004.tex

\section{INTRODUCTION}
\label{sec:Motivations}
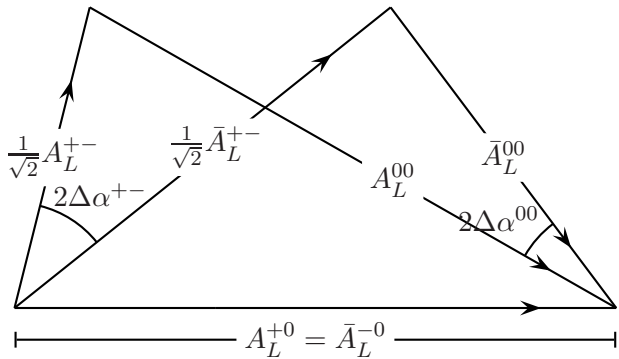
\begin{figure}[htbp]
  \begin{center}
    \begin{pspicture}(0.7,-0.8)(9.1,4.1)
      \pnode(1,0){zero}
      \pnode(9,0){one}
      \pnode(3.66667,0){a2}
      \pnode(2,4){bz}
      \pnode(6,4){bzb}
      \ncline[offset=-12pt]{|-|}{zero}{one}
      \lput*{0}{$A^{+0}_{L} = \bar{A}^{-0}_{L}$}
      \ncline{zero}{a2}
      \ncline{a2}{one}
      \ncline[linestyle=none,nodesep=1,arrowsize=.2]{->}{a2}{one}
      \ncline{zero}{bz}
      \ncline[linestyle=none,nodesep=1,arrowsize=.2]{->}{zero}{bz}
      \lput*{0}(0.5){$\frac{1}{\sqrt{2}}A^{+-}_{L}$}
      \ncline{zero}{bzb}
      \ncline[linestyle=none,nodesep=1,arrowsize=.2]{->}{zero}{bzb}
      \lput*{0}(0.55){$\frac{1}{\sqrt{2}}\bar{A}^{+-}_{L}$}
      \ncline{bz}{one}
      \ncline[linestyle=none,nodesep=1,arrowsize=.2]{->}{bz}{one}
      \lput*{0}(0.6){$A^{00}_{L}$}
      \ncline{bzb}{one}
      \ncline[linestyle=none,nodesep=1,arrowsize=.2]{->}{bzb}{one}
      \lput*{0}{$\bar{A}^{00}_{L}$}

   \SpecialCoor
      \psarcn(zero){1.4}{(1,4)}{(5,4)}
      \uput{1.45}[60](zero){$2\Delta\alpha^{+-}$}

     \psarcn(one){1.4}{(-7,4)}{(-3,4)}
      \uput{1.45}[135](one){$2\Delta\alpha^{00}$}

    \end{pspicture}
    \caption{Isospin Triangles relating the amplitudes of the modes 
$\Bz \to \rhop\rhom$, $\Bz \to \rhoz\rhoz$, $\Bp \to \rhop\rhoz$
and their charge conjugates for a given polarization
(e.g., \CP-even longitudinal polarization).}
    \label{fig:triangle}
  \end{center}
\end{figure}

Measurements of \CP--violating asymmetries in the \BzBzb system allow
tests of the Standard Model by over-constraining the Unitarity
Triangle through the measurement of its angles. 
The time--dependent \CP asymmetry in a $\b \to \u{\bar \u}\d$ decay 
of a \Bz to a \CP 
eigenstate allows for a direct measurement of the angle 
$\alpha\equiv \arg\left[-V_{td}^{}V_{tb}^{*}/V_{ud}^{}V_{ub}^{*}\right] $  
if the decay is dominated by the tree amplitude. The 
contribution from penguin diagrams gives rise to a correction 
$\Delta\alpha= \alpha_{\rm eff} - \alpha $ that can be inferred through an
isospin analysis~\cite{gronau90}  
as illustrated in Fig.~\ref{fig:triangle}. The theoretical error
from isospin--breaking effects is typically estimated to be smaller
than 10 degrees~\cite{ciuchini}.

The $\pip\pim$ final state (\CP=1) was expected to be an ideal candidate for 
the study of $\alpha$. However, the recent measurement of 
$\Bz \to \pi^0\pi^0$ confirms that this mode suffers from substantial
penguin contamination~\cite{pi0pi0}. This severely limits the possibility of
a model--independent
measurement of $\alpha$ with currently available statistics.
On the other hand, the recent experimental results shown in 
Table~\ref{tab:stateofart}
confirm the theoretical expectation~\cite{aleksan} of a 
smaller penguin contribution 
in the $\rho\rho$ system
with respect to the $\pi\pi$ case (charge-conjugate decay modes are always 
assumed throughout this paper).
\begin{table}[htb]
\begin{center}
\caption{Recent measurements of $B\to\rho\rho$ branching fractions.} 
\vspace{0.2cm}
\begin{tabular}{lcc}  
\hline
Decay                                           &  \babar   &    Belle     \\
\hline
        &    &    \vspace*{-0.3cm} \\                                       
$BR(B^+ \to \rho^+\rho^0)\times 10^6$           & 
22.5$^{+5.7}_{-5.4}\pm$5.8~\cite{Aubert:2003mm} &
31.7$\pm$7.1$^{+3.8}_{-6.7}$~\cite{Zhang:2003up}  \\
        &    &     \vspace*{-0.3cm} \\                                       
$BR(B^0 \to \rho^+\rho^-)\times 10^6$           &
30$\pm 4\pm 5$~\cite{Aubert:2003xc}             &
                                                   \\
        &    &     \vspace*{-0.3cm} \\                                       
$BR(B^0 \to \rho^0\rho^0)\times 10^6$           &
$<$2.1 (90 \% C.L.)~\cite{Aubert:2003mm}        &  
                                                   \\
        &    &     \vspace*{-0.3cm} \\                                       
\hline
\end{tabular}
\label{tab:stateofart}
\end{center}
\end{table}

In $\rho\rho$ decays the final state is not, in general, a \CP
eigenstate, and an isospin triangular relation holds for each of the
three helicity states, which can be separated through an angular
analysis. However, the measured polarizations in $\rho^+\rho^-$ and
$\rho^+\rho^0$ modes indicate a dominance of the helicity 0 state (longitudinal
polarization), that is, of a \CP=+1 eigenstate. 
A measurement of the polarization in $\Bz \to \rhoz\rhoz$
would complete the isospin triangle, but this mode has not
been observed so far.
The best present limit on the \rhozrhoz decay was obtained by \babar\ with a
sample of 89 million \upsbb decays~\cite{Aubert:2003mm}. 
The resulting Grossman--Quinn~\cite{grossman98} bound
that can be set on the penguin contribution~\cite{Aubert:2003xc, falketal}
is more stringent than in the $\pi\pi$ case, but 
knowledge of the $\Bz \to \rhoz\rhoz$ rate is still expected
to be the limiting factor
to the accuracy of the $\alpha$ measurement 
with $\rho\rho$ decays. In~\cite{grossman98}, Grossman and Quinn show
that the angle $2\Delta\alpha^{+-}$  (see Fig.~\ref{fig:triangle})
between the two triangles is maximum 
when the two isospin triangles have opposite direction and they are
right triangles. In that case, the bound takes simply the form
\begin{equation} \label{eq:grossquinn}
\sin^2\Delta\alpha^{+-} \leq
\frac
{f^{00}_L\times {\cal B}(\Bz \to \rhoz\rhoz)}
{f^{+0}_L\times {\cal B}(\Bp \to \rhop\rhoz)}
\end{equation}
where 
$f^{00}_L$ and $f^{+0}_L$ are the fraction of longitudinal polarization in
$\Bz \to \rhoz\rhoz$ and $\Bp \to \rhop\rhoz$ decays respectively, and
${\cal B}(\Bz \to \rhoz\rhoz)$ and ${\cal B}(\Bp \to \rhop\rhoz)$ the
respective branching fractions.

In this paper, we present a search for the \rhozrhoz final state
performed at \babar\ on a sample of 227 million \upsbb decays. 
Improvements over the previous result are achieved from 
the increased statistics and from optimizations 
of the analysis, which result in an increased sensitivity.

\section{THE \babar\ DETECTOR AND DATASET}
\label{sec:Detector}

The \babar\ detector operates at the \pep2\ \abf\ at SLAC and is
described in detail elsewhere~\cite{babar}. The data samples used in
this analysis correspond to an integrated luminosity of 205.4~\invfb
collected at the \FourS resonance, and of 16.1~\invfb collected 
about 40~\mev below the~\FourS resonance in order to study the \qqbar 
continuum background. 
Large samples of Monte Carlo data are also used to model signal
$B$ backgrounds.

\section{ANALYSIS METHOD}
\label{sec:Analysis}

We fully reconstruct $B^0\to\rho^0\rho^0$ candidates from 
their decay products $\rho^0\to \pi^+\pi^-$ with four
charged tracks in the final state. Charged track candidates 
are required to originate from a single vertex near the 
interaction point. We select $\rho^0$ candidates 
with requirements on the $\pi^+\pi^-$ invariant mass,
loose enough to retain sidebands for later fitting.

Improvements in the track
reconstruction efficiency and in the understanding of the associated
systematic error have allowed us to extend the accepted momentum range with
respect to the previous analysis, and consequently increase the selection
efficiency by 25\%. The particle identification
capabilities of the \babar\ detector are used to reject tracks
identified as electrons, kaons, or protons. A set of kinematic
variables describing the shape of the event is used to suppress the
\qqbar continuum background. The identification of signal
events is based on two kinematic variables: 
the beam--energy substituted mass of the \B 
\begin{equation} \label{eq:mes}
\mes = \sqrt{(s/2 + {\mathbf {p}}_i\cdot {\mathbf{p}}_B)^2/E_i^2-
{\mathbf {p}}_B^2}
\end{equation}
(where the initial four--momentum $(E_i, {\mathbf {p_i}})$ and the \B
momentum ${\mathbf {p_B}}$ are defined in the laboratory frame), and
the difference between the reconstructed \B energy in the
center--of--mass frame and its known value
\begin{equation} \label{eq:deltae}
\DeltaE = E_B^{\rm CM} - \sqrt{s}/2.
\end{equation}
An unbinned extended maximum likelihood fit is then
performed on the selected sample to evaluate the signal yield 
using several discriminating variables. 
The probability density functions (PDF) entering the 
likelihood function  are built as the sum of several components that
describe separately the signal, the \qqbar continuum, and
various background \BB decays. 
The result was revealed only after the finalization and validation
of the full analysis.

\subsection{Angular Observables}
\label{sec:angular}

\begin{figure}[htbp]
\begin{center}
\centerline{
\epsfig{figure=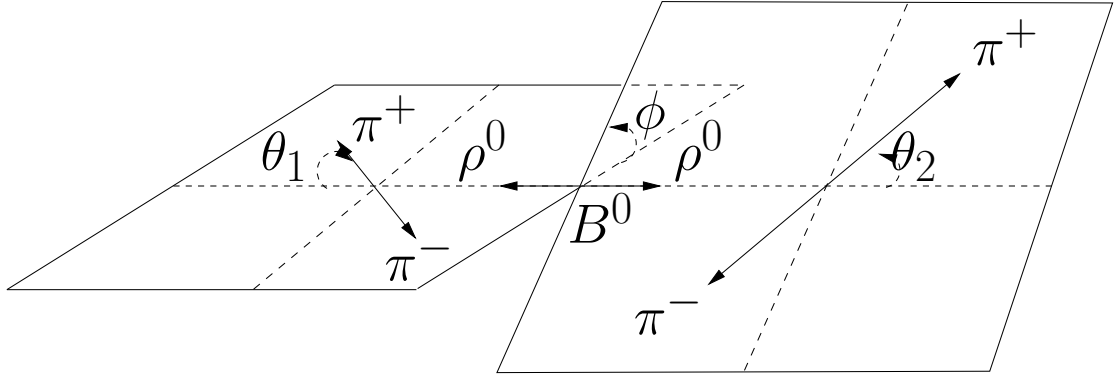,width=6.0in}
}
\caption{ 
Definition of helicity angles $\theta_1$, $\theta_2$, and
$\phi$, for the decay ${B^0\to\rho^0\rho^0}$. The $\rho^0$
final states are shown in their rest frames.
}
\label{fig:helangles}
\end{center}
\end{figure}

The angular distribution of the $B$ meson decay to a
vector-vector ($VV$) final state is a combination of S-,
P-, and D-wave contributions with unknown relative amplitudes.
The helicity angles 
($\theta_i$, ${\phi}_i$, $\phi=\phi_1-\phi_2$, $i=1,2$) 
are defined by the direction of the two-body $\rho^0$ decay 
axis and the direction opposite the \B in the $\rho^0$ rest system, 
as shown in Fig.~\ref{fig:helangles}.

Due to quark spin conservation, the longitudinal polarization  
$A_L$ is expected to dominate the $VV$ decays.
However, this expectation might be strongly violated
in penguin--dominated decays, as has been observed in
$B\to\phi K^*$ ~\cite{Aubert:2003mm}.
Since the azimuthal angle $\phi$ does not provide any significant background 
suppression, and there are no acceptance effects for this observable, 
we integrate over it, which results in:

\begin{equation} \label{eq:helicityshort}
{1 \over \Gamma} \ 
{d^2\Gamma \over d\cos \theta_1~d\cos \theta_2} = {9 \over 8\pi} 
\left[
 \cos^2 \theta_1 \cos^2 \theta_2 f_L +
{1 \over 4} \sin^2 \theta_1 \sin^2 \theta_2  
\left( 1 - f_L \right)  \right]
\end{equation}

The two helicity observables ${\cal H}_i=|\cos\theta_i|$ are used in
the fit to measure the longitudinal polarization
of the decay, $f_L=|A_0|^2/(|A_0|^2+ |A_{+1}|^2 + |A_{-1}|^2)$,
where $A_\lambda\, (\lambda = 0, +1,-1)$ correspond to the helicity
amplitudes of the two $\rho$'s in the $B$ rest system.
These observables also provide additional background suppression.
The limit on the branching fraction can be expressed as a function 
of $f_L$. 
The hypothesis $f_L=1$ gives the most conservative upper limit
on the $A^{00}_L$ amplitude, shown in Fig.~\ref{fig:triangle}.
This is also the relevant result for the measurement of $\alpha$ through the
isospin analysis. 

\subsection{Event Selection}
\label{sec:selection}

In order to reject the dominant quark-antiquark continuum background,
we require $|\cos\theta_T| < 0.8$, where $\theta_T$ 
is the angle between the $B$-candidate thrust axis
and that of the remaining tracks and neutral clusters in
the event, calculated in the CM frame. 
The other event-shape discriminating variables include the polar 
angles of the $B$ momentum vector and the $B$-candidate thrust 
axis with respect to the beam axis in the $\Upsilon(4S)$ frame, 
and the two Legendre moments $L_0$ and $L_2$ of the energy 
flow around the $B$-candidate thrust axis.
These variables are combined in a neural network, which is trained 
using off-resonance beam data and signal Monte Carlo data.
The neural network output is transformed into a Gaussian-shaped
variable (called ${\cal E}$-variable hereafter), without any loss of 
discriminant power, to make the shape easier to fit.

To further suppress background, 
we use multivariate algorithms to identify the flavor
of the other $B$ in the event (tagging)~\cite{babarsin2beta}.
The suppression power comes from the fact that signal and
background have different tagging efficiencies in the
various tagging categories.
These categories correspond to different methods of 
identifying the \b content of the other \B meson in the event. 
We use five tagging categories $c_{\rm tag}$ (lepton, 
two kaon, inclusive, and the untagged event categories).
Since \BB events tagged by the best performing tagging categories 
(notably by leptons) have a lower \qqbar background, the splitting 
into tagging categories improves the discrimination of the continuum. 

To summarize, we use eight primary observables to
characterize each event candidate: 
$\vec{x}_{i}=
\{m_{\rm{ES}}$, 
$\Delta E$, 
${\cal E}$, 
$m(\pi^+\pi^-)_1$, $m(\pi^+\pi^-)_{2}$, 
${\cal H}_1$, ${\cal H}_2$, $c_{\rm tag}\}$.
The ranges for the key discriminating variables are the following:
we require $\mes > 5.24$ Gev/c$^2$ and $|\DeltaE|<85~\mev$ 
(the signal $m_{\rm{ES}}$ and $\DeltaE$ resolutions are
2.5 MeV/c$^2$ and 20~\mev respectively), 
$m(\pi^+\pi^-)_{i}$
between 0.55 and 1.0 GeV/c$^2$,  and ${\cal H}_i<0.99$. The latter
cut removes a region with very low efficiency.

Additional selection criteria are finally applied to veto the most dangerous 
source of \BB to charm background, namely 
$\Bz\to\Dm\pip\to \Kp (\pip) \pim\pim\pip$, by requiring the 
invariant mass of the three--particle combination that excludes the
highest--momentum track in the \B frame to be inconsistent with 
a $D$ meson.

After applying all selection criteria, the event sample has on average
1.05 candidates per event. When several candidates are selected in the
same event, one candidate is selected randomly.
The selection efficiency, measured on signal Monte Carlo samples,
is 27\% (32 \%) for the longitudinaly (transversely) polarized
signal. This corresponds to an expected signal yield in our data sample
of about 60 events for a branching fraction of $1\times 10^{-6}$.
The fraction of selected events where the four candidate
tracks do not match the true event tracks, called self--cross--feed
(SCF) events in the following, is 22\% (8\%) for longitudinal
(transverse) polarization. The large self--cross-feed for longitudinal
signal is due to the presence of soft pions that can
be easily exchanged with soft particles from the other \B.

\clearpage

\subsection{Backgrounds}
\label{sec:backgrounds}

The selected sample is expected to be dominated by \qqbar
combinatorial background. However, the most dangerous backgrounds are
expected to come from other \B decays, in particular from final
states with four charged particles that can mimic the signal in the
distribution of $\mes$ and $\DeltaE$, for four--pion final states.
As mentioned above, the \B to charm decays are
suppressed by explicit veto cuts. For the \B to charmless decays, we expect
contaminations from more than 50 modes, the most important ones being
listed in Table~\ref{tabBbkg}. The mode 
$B^0\to a_1^{\pm}\pi^{\mp}\to\rhoz\pi^{\pm}\pi^{\mp}$ 
is an irreducible source of background.
The discrimination of this mode relies mainly on the $\rho$ mass
distributions. 

The other charmless modes have smaller impact on the result.
The most important ones are
described by specific PDFs, in order to improve the background model in
the likelihood fit.
We fix the yields of well measured modes ($\rho^+\rho^0$,
$\rho^+\rho^-$, $\rho\pi$) and modes which are isospin-related 
to measured modes ($\rho^0K^{*0}$), and vary these yields to
evaluate the corresponding systematic error.

\begin{table}[htb]
  \centering
  \caption{Background categories which are either floated or
fixed in the fit to the data sample. 
The yield uncertainties in the fixed background categories 
(both statistical and systematic)
are taken into account in the evaluation of the systematic error.
}
\vspace{0.2cm}
  \begin{tabular}{lc}  
\hline
 Background Category   &  Yield \\ \hline 
 $q\bar{q}$                             & floated  \\
 $\B\to\text{charm}$                    & floated  \\
 $B^0\to a_1^{\pm}\pi^{\mp}$            & floated  \\
 $B^0\to \rho^0K^{*0}$                  & 25 $\pm$ 5 $\pm$ 16 ~\cite{Aubert:2003mm} \\  
 $B^+\to \rho^+\rho^0$                  & 78 $\pm$ 9 $\pm$ 18 ~\cite{Aubert:2003mm} \\  
 $B\to\rho\pi$ modes                    & 34 $\pm$ 6 $\pm$  4 ~\cite{HFAG}  \\  
 $B^0\to\rho^+\rho^-$                   & 14 $\pm$ 4 $\pm$  3 ~\cite{Aubert:2003xc} \\  
 other charmless $B$ decays                  & floated \\  
    \hline
  \end{tabular}
  \label{tabBbkg}
\end{table}

\subsection{Likelihood Fit}
\label{sec:fit}

The signal is obtained by maximizing the extended likelihood function 
\begin{equation}
{\cal L} = \exp\left(-\sum_{j=1}^{N_{\rm cat}} n_{j}\right)\, 
\prod_{i=1}^{N_{\rm cand}} 
\left(\sum_{j}~n_{j}\, 
{\cal P}_{j}(\vec{x}_{i};\vec{\beta})\right)
\label{eq:likel}
\end{equation}
with several event categories $j$: signal (including 
a SCF fraction), continuum~$\qqbar$, and several 
$\BB$ background categories listed in Table~\ref{tabBbkg}.
Each ${\cal P}_{j}(\vec{x}_{i};\vec{\beta})$ is 
the PDF for the observable
$\vec{x}_{i}$,
and is described by the PDF parameters $\vec{\beta}$. 

The event yields $n_j$ in different categories
are obtained by minimizing the quantity $-2\ln{\cal L}$. 
The statistical significance of a signal is defined as the 
square root of the change in $-2\ln{\cal L}$ when constraining 
the number of signal events to zero in the likelihood fit.
The assumption of negligible correlations among most of the fit 
input variables has been validated with MC simulation and data.
However, there are effects which do introduce noticeable
correlations in the samples of signal and background events, such as
helicity angles correlation in signal, and
mass-helicity correlation in background, as discussed below.

\clearpage

\subsection{PDF Parameterization}
\label{sec:pdf}

We use double-Gaussian functions to parameterize the 
$\Delta E$ and $m_{\rm{ES}}$ PDF's for signal,
and a relativistic P-wave Breit-Wigner 
convoluted with a Gaussian resolution function
for the resonance masses.
The ${\cal E}$-variable is described by an asymmetric 
Gaussian plus a single Gaussian function for both signal 
and background.
The signal helicity PDF is expressed as a function of the 
longitudinal polarization (see Eq.~\ref{eq:helicityshort}).
The ideal angular distribution is multiplied by the detector 
acceptance function ${\cal{G}}({\cal H}_1,{\cal H}_2)$. 
We obtain the acceptance function from a fit 
to a sample of MC events with transverse and longitudinal 
polarization.

\begin{figure}[htbp]
\begin{center}
\begin{tabular}{cc}
\epsfig{figure=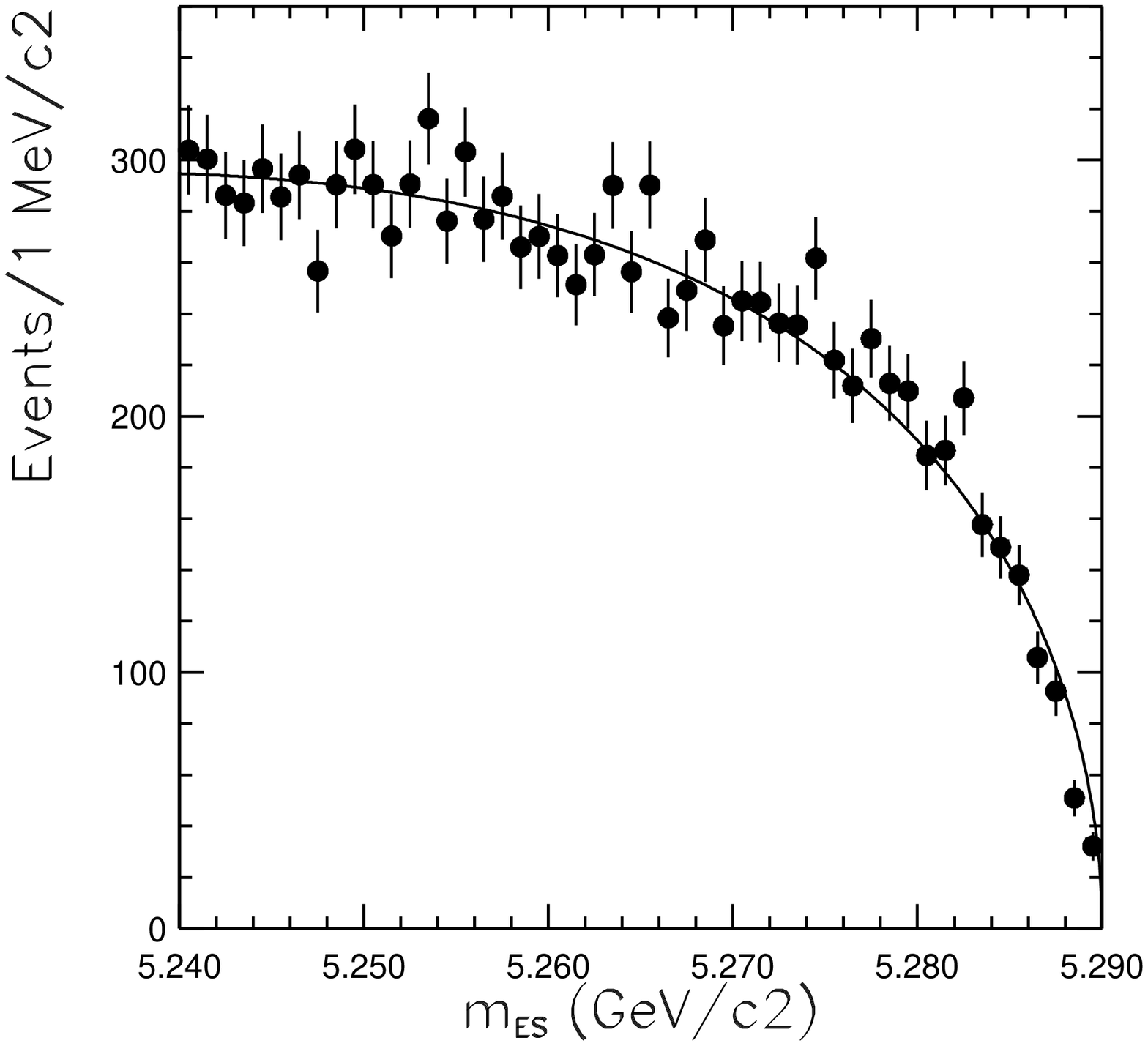,height=1.9in} &
\epsfig{figure=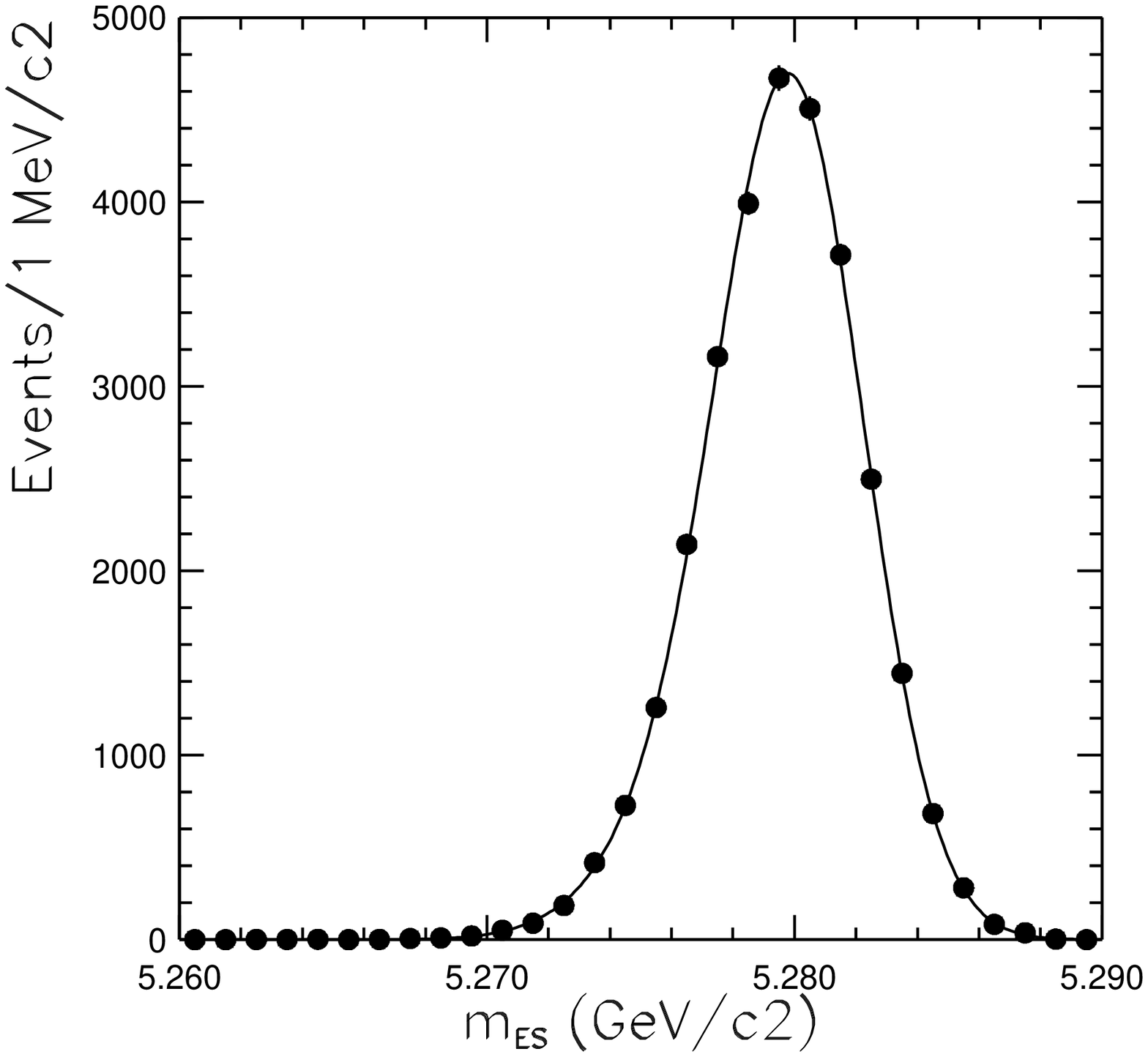,height=1.9in} \\
\epsfig{figure=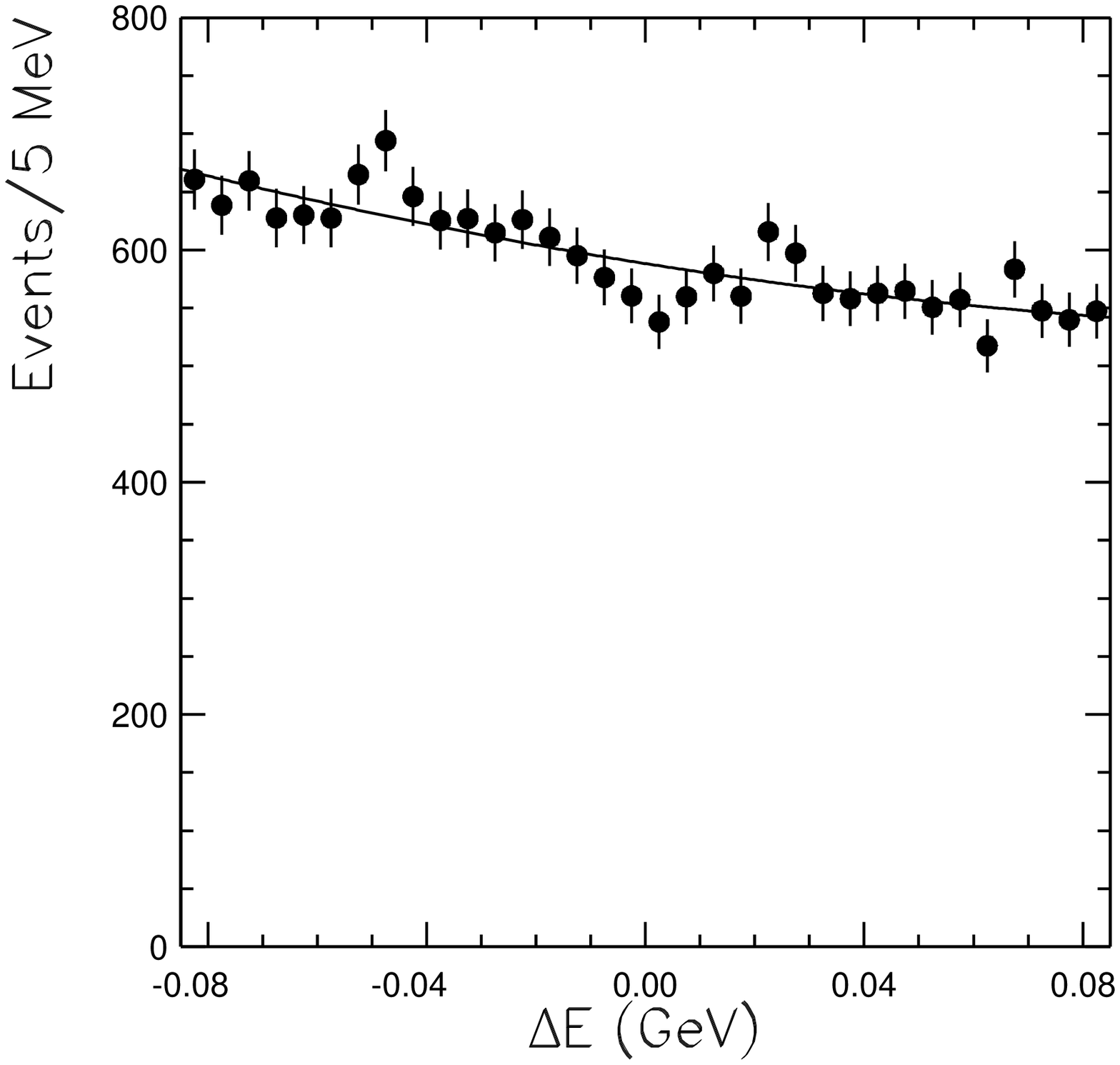,height=1.9in} &
\epsfig{figure=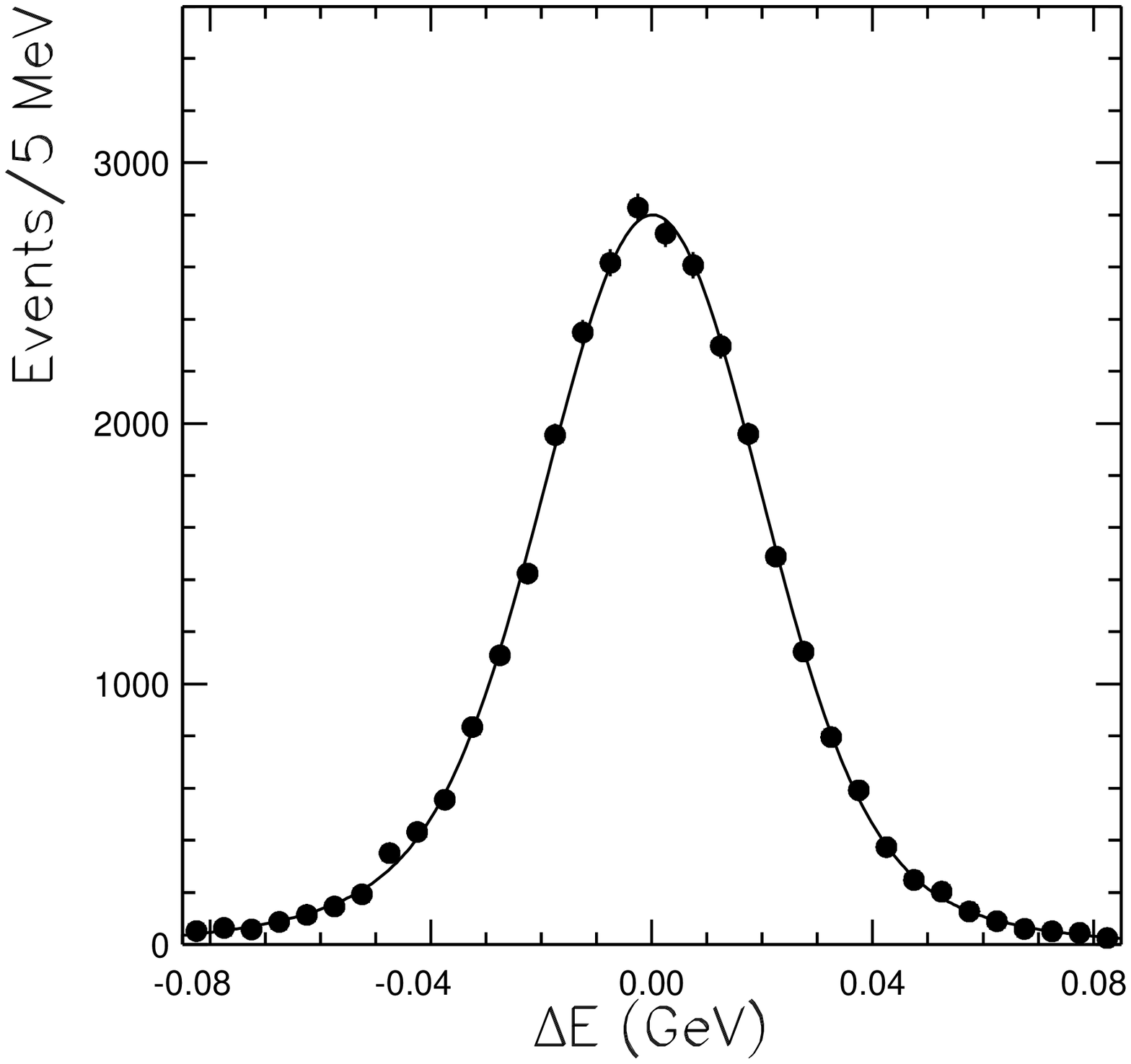,height=1.9in} \\
\epsfig{figure=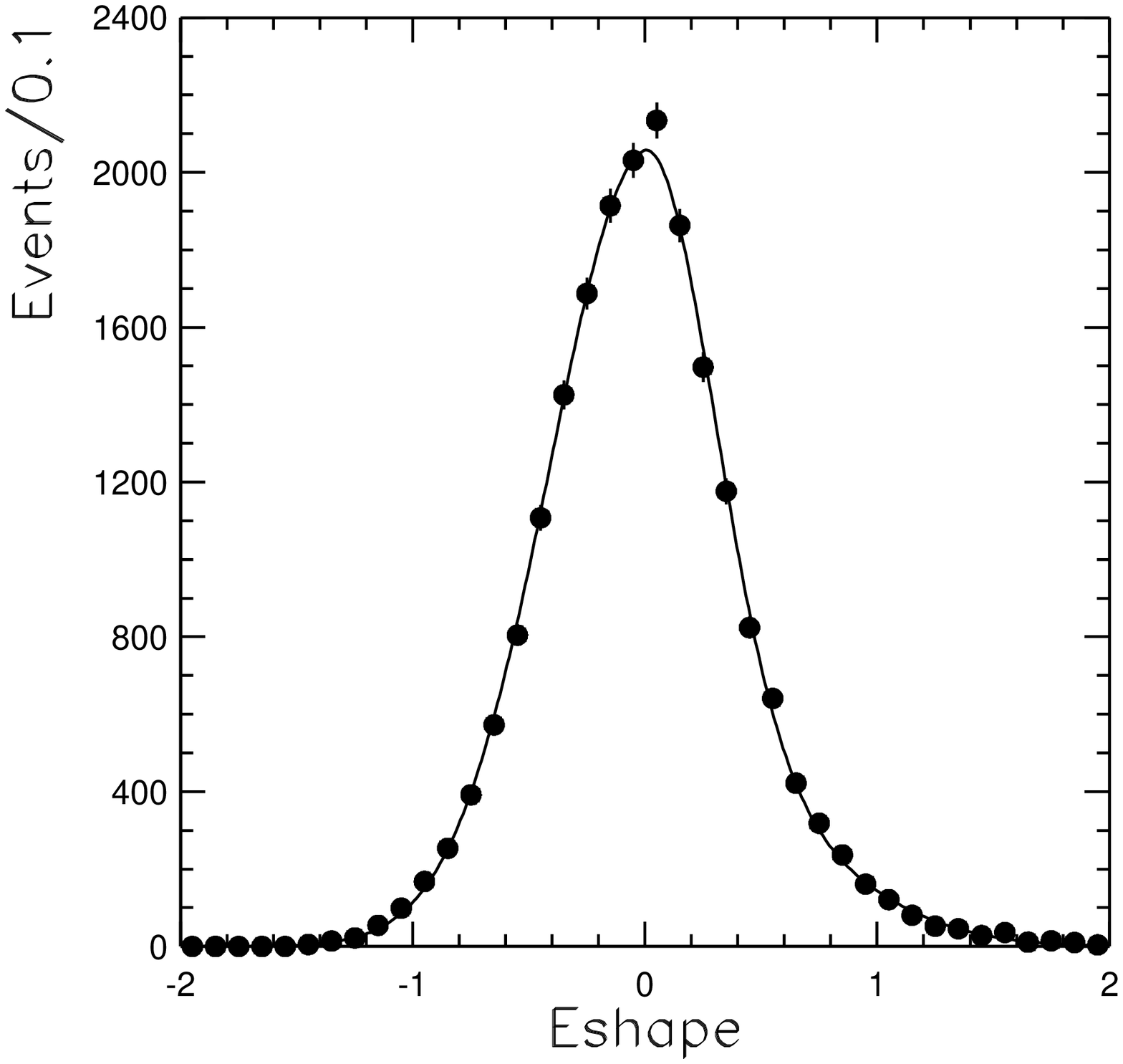,height=1.9in} &
\epsfig{figure=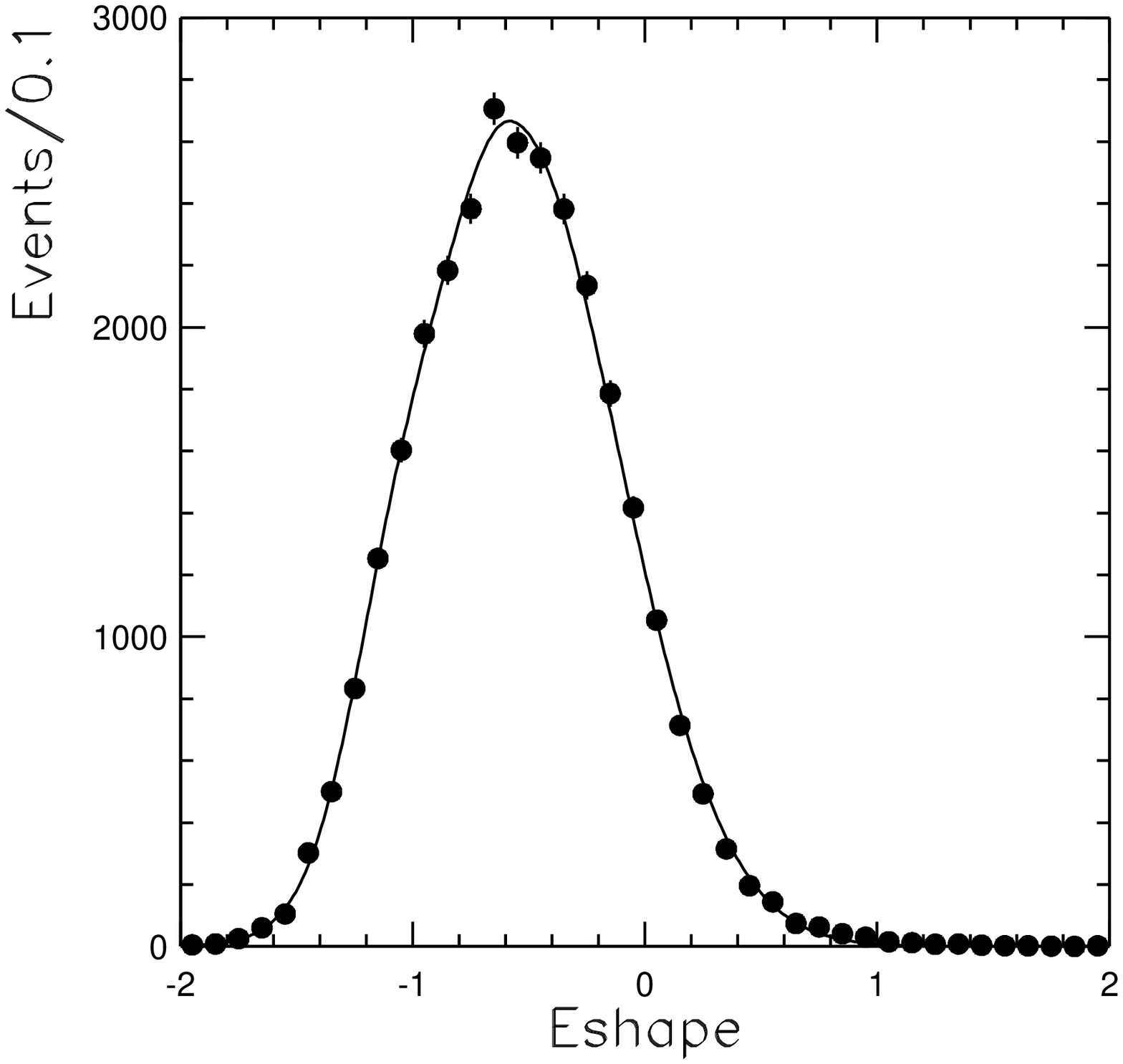,height=1.9in} \\
\epsfig{figure=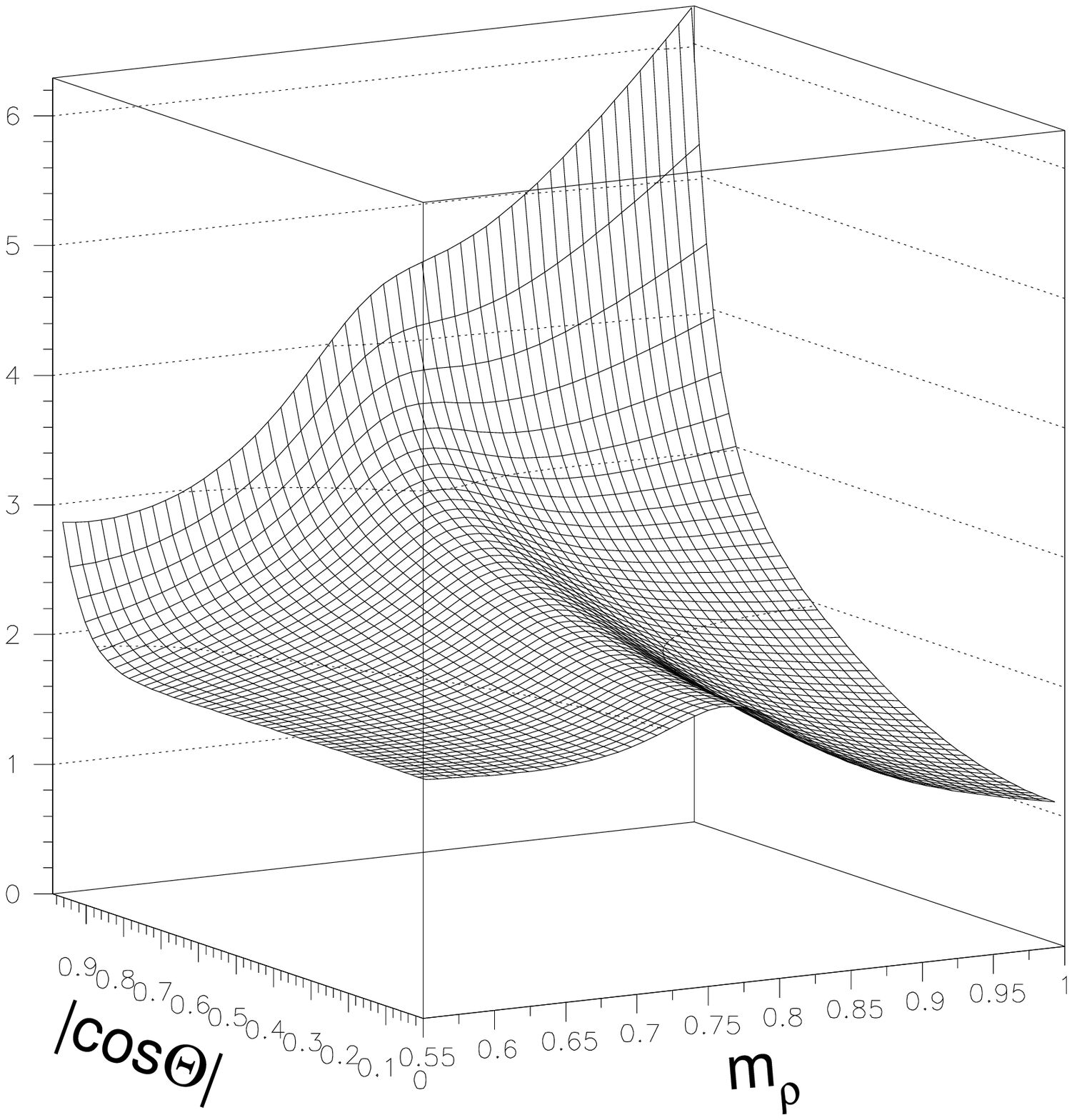,height=1.9in} &
\epsfig{figure=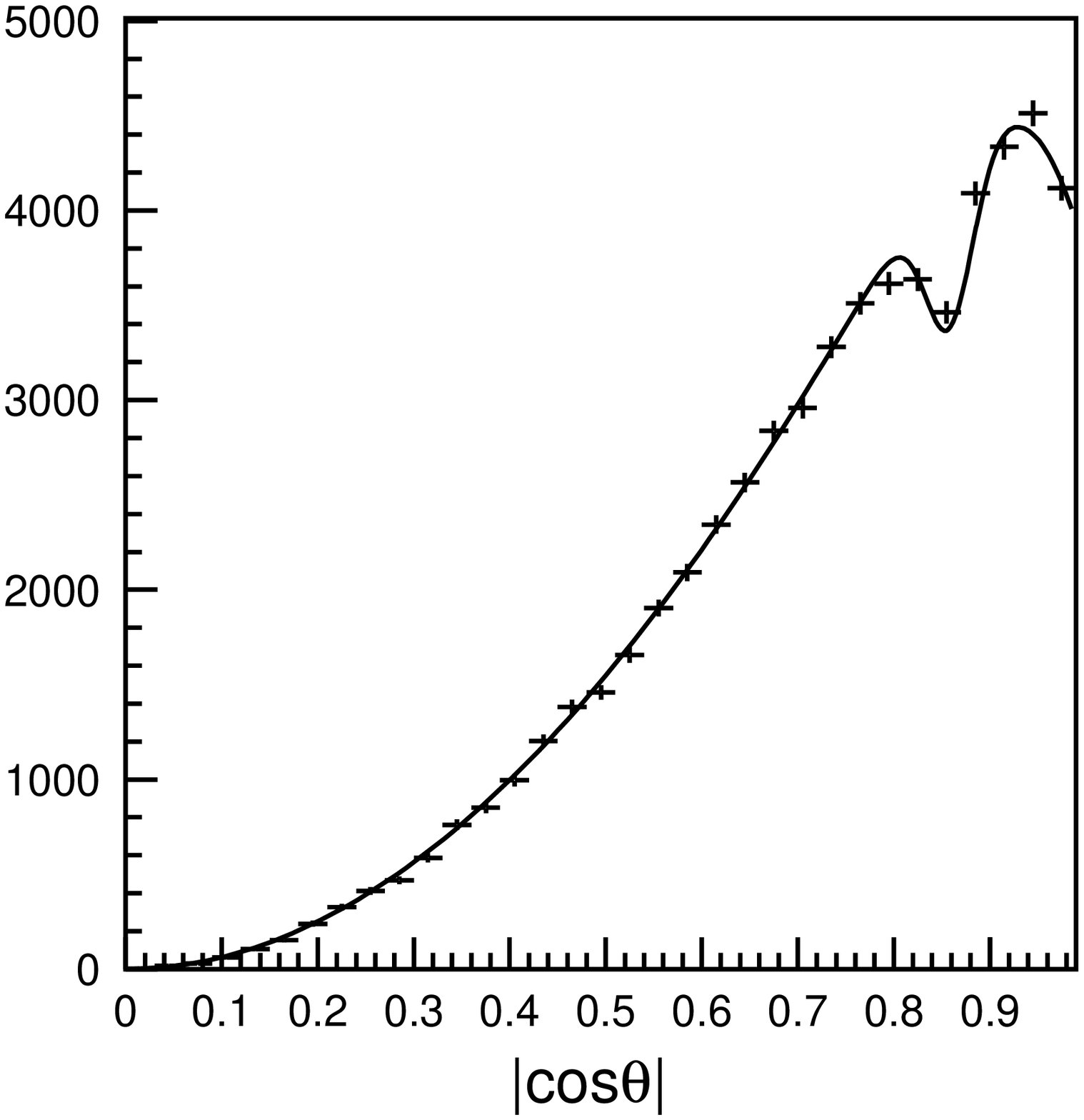,height=1.9in} \\
\end{tabular}
\caption{PDFs of the observables (from top to bottom):
$m_{\rm{ES}}$, $\Delta E$, and ${\cal E}$, for
combinatorial background (left) and longitudinal signal (right).
The bottom row shows the combinatorial background
mass-helicity two-dimensional PDF (left), and 
the projection of the helicity distribution for signal (right).
The small dip near $|\cos\theta| = 0.85$ is due to the $D$-veto acceptance
effect.}
\label{fig:pdfs}
\end{center}
\end{figure}

For the combinatorial background we use low-degree polynomials 
for $\Delta E$ and resonance masses,
and an empirical phase-space function for $m_{\rm{ES}}$
(known as Argus function~\cite{argus}):
\begin{eqnarray}
f(x) \propto x\sqrt{1-x^2}~{\rm exp}[-\xi(1-x^2)] \ ,
\label{eq:bmass}
\end{eqnarray}
where $x = m_{\rm{ES}}/E_{\rm beam}$, 
and $\xi$ is a parameter measuring the curvature of the distribution
near the end-point.
The background parameterization of resonance masses also 
includes a resonant component to account for resonance production,
which is described by 
the same P-wave Breit-Wigner function used for the signal $\rho$ shape.
The background helicity-angle distribution is 
separated into contributions from combinatorial background
and from resonances. The resonances are assumed to
be unpolarized.
We parameterize the combinatorial helicity distribution
with a second-degree polynomial and an exponential
function to allow for the increased fraction of fake $\rho$
candidates with low momentum pions near ${\cal H}_i=1$.
The amount of peaking near ${\cal H}_i=1$ 
depends on the $\rho$ candidate mass and is parameterized
accordingly. 
Fig.~\ref{fig:pdfs} shows the $m_{\rm{ES}}$, $\Delta E$, and ${\cal E}$
PDFs for combinatorial background and longitudinal signal.
The two-dimensional mass-helicity distribution for combinatorial
background, and the projection of the helicity distribution for
longitudinal signal are also shown.

The PDFs for exclusive \B decay modes are modeled with 
non-parametric distributions describing the shapes of the
observables, except when certain distributions are expected
to be identical to those used for signal. These distributions are described 
by smoothed histograms~\cite{nonparam} with a large number of bins. 
In the $B$-background modes the two $\rho^0$ candidates can 
have very different mass and helicity distributions. This happens, 
for example,
when one of the two $\rho^0$s is real (e.g. $\rho^+\rho^0$, 
$\rho^0K^{*0}$) or when only one of the two $\rho^0$s contains a hard 
bachelor pion ($a_1\pi$). In such cases, we consider a 
four-variable correlated mass/helicity PDF.
The inclusive $B$-to-charm and $B$-to-charmless categories
are parameterized similarly to the exclusive modes, they
include the remaining $B$ decay modes not modeled explicitely,
and their yield is left free in the fit, to include 
possible unaccounted background sources and compensate 
for possible imperfections in the background model.

The $B$-flavor tagging PDFs for signal and background are
simply the discrete distributions of tagging efficiencies.
Large samples of fully reconstructed $B$ meson decays are
used to obtain the $B$-tagging efficiencies for signal $B$ decays 
and to control the MC values of $B$-tagging efficiencies for the 
\B backgrounds. Continuum background efficiencies are obtained 
from the sideband data.

\clearpage

\section{PHYSICS RESULTS}
\label{sec:Physics}
Table~\ref{tab:results} shows the results of the fit. No significant
yield is observed, and an upper limit on the branching fraction
of \Btozz is set. As mentioned earlier, 100\% longitudinally polarized
signal is used in the fit, as it was checked that it gives the most 
conservative estimate for the upper limit.

\begin{table}[htb]
  \centering
  \caption{ 
Results of the fit: Signal yield ($N_S$), selection
efficiency (Eff), branching fraction (${\cal B}$),
upper limit at 90\% confidence level (U.L.) and significance
of the measurement, defined as 
$\sqrt{\Delta (-2\ln{\cal L})}$
when constraining the number of
signal events to zero in the fit.
The first error corresponds to the statistical
uncertainty and the second one to the systematic uncertainty,
discussed in next section.
}
\vspace{0.2cm}
  \begin{tabular}{lc}  
\hline
Quantity                      &  Value             \\
\hline                     
                              &   \vspace*{-0.3cm} \\
$N_S$                         &   $33^{+22}_{-20}\pm 12$ \\
                              &   \vspace*{-0.3cm} \\
Eff (\%)                      &  $27.1\pm 1.3$       \\
                              &   \vspace*{-0.3cm} \\
${\cal B}(\times 10^{-6})$    &   $0.54^{+0.36}_{-0.32}\pm 0.19$ \\
                              &   \vspace*{-0.3cm} \\
U.L.$(\times 10^{-6})$        &   $1.1$ ($1.0$ statistical only) \\          
                              &   \vspace*{-0.3cm} \\
Significance ($\sqrt{\Delta (-2\ln{\cal L})}$)       &   
                              $   1.6$ ($1.9$ statistical only) \\     
                              &   \vspace*{-0.4cm} \\
\hline
  \end{tabular}
  \label{tab:results}
\end{table}

\begin{figure}[htb]
\begin{center}
\begin{tabular}{cc}
  \epsfig{figure=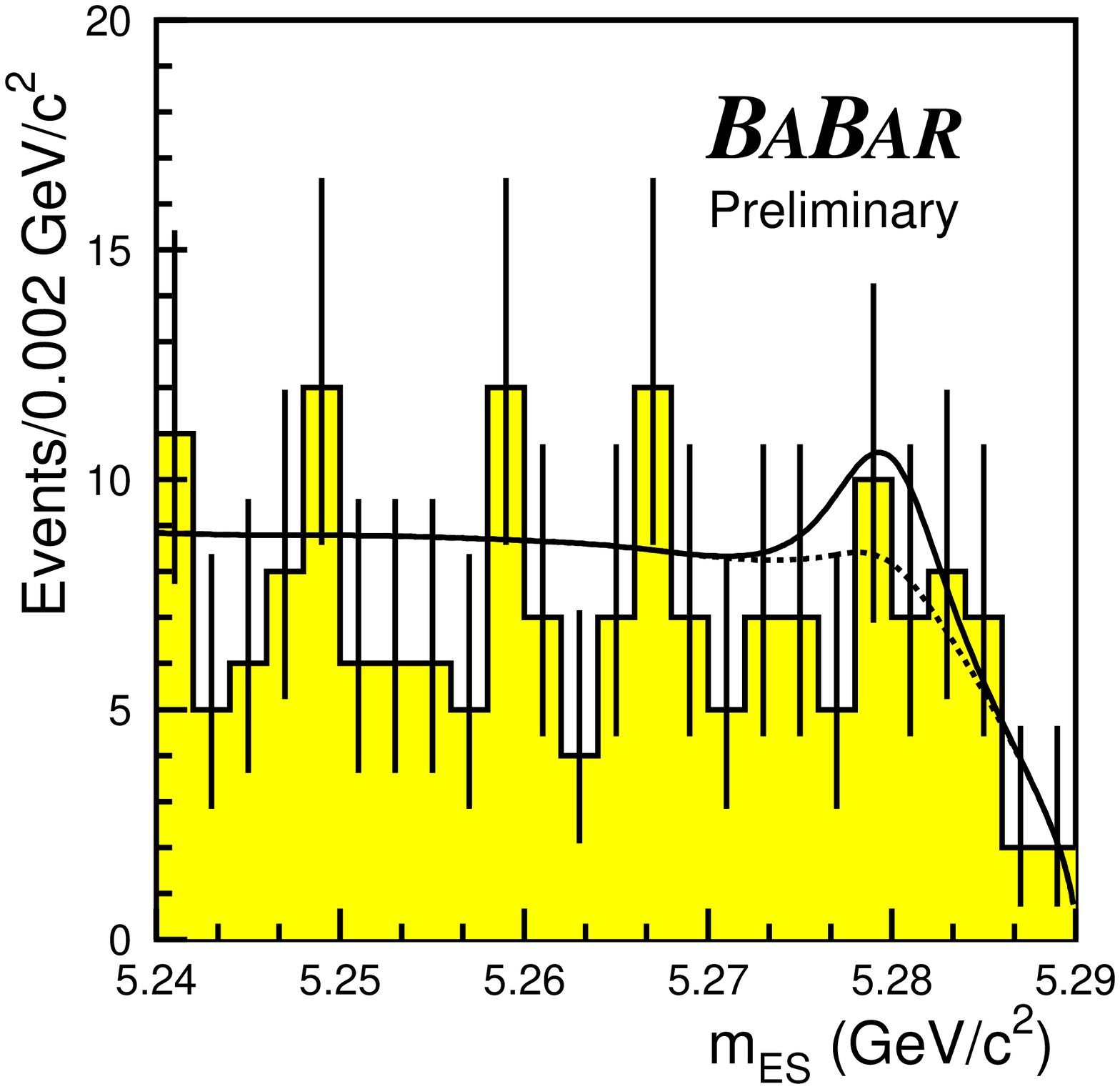,height=.4\textwidth}  &
  \epsfig{figure=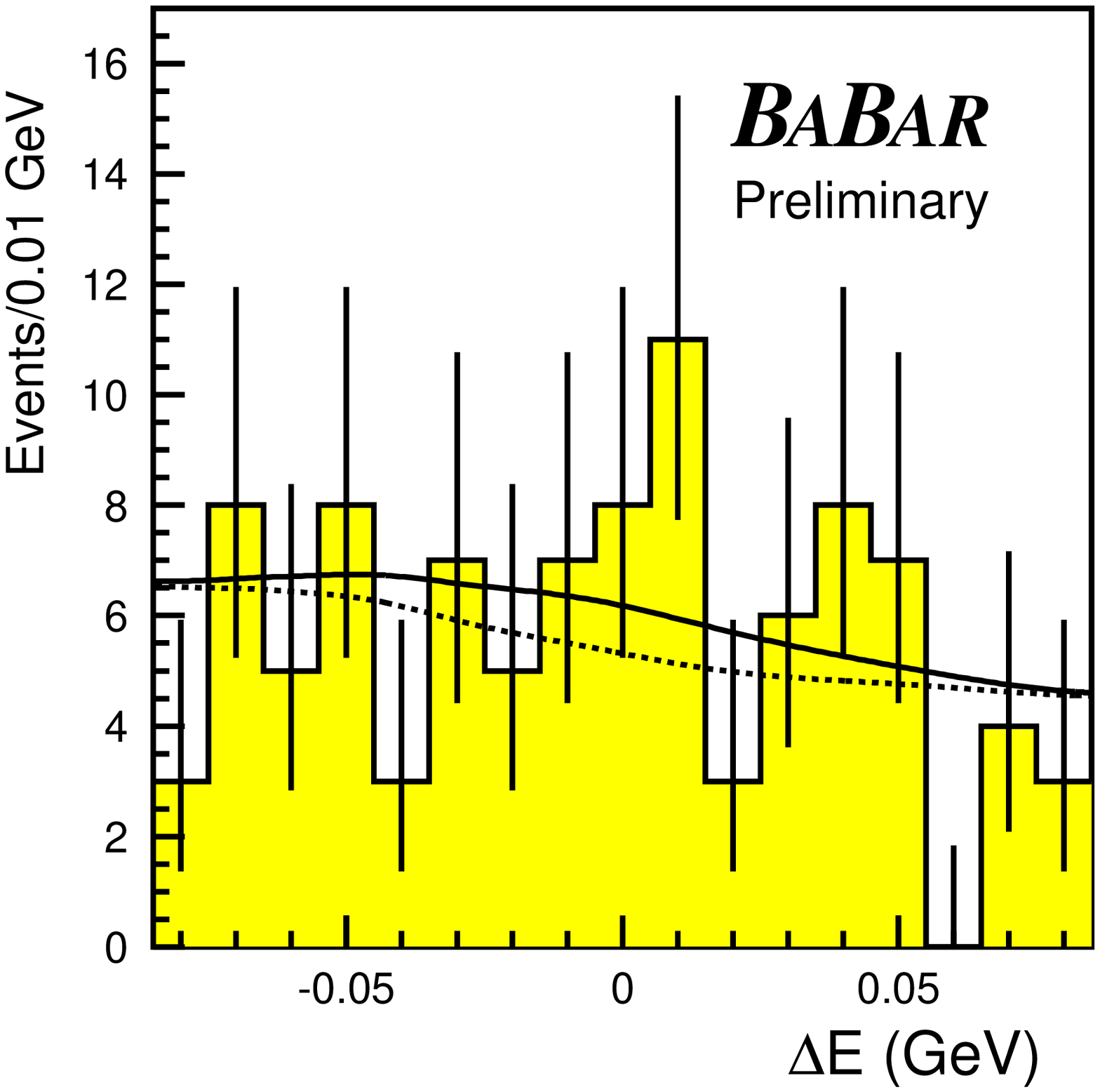,height=.4\textwidth}  \\
\end{tabular}
\label{fig:projections}
\end{center}
\caption{Fit result projections onto $m_{ES}$ and $\Delta E$. 
The histograms correspond to the data and the solid (dashed) line to the 
full (background only) fit after a cut on the
quantity ${\cal P}_{\rm sig}/({\cal P}_{\rm sig}+{\cal P}_{\rm back})$ 
is applied.
The projections contain 22.5\% and 23.9\% of signal, and less than
0.5\% and 0.2\% of continuum background respectively.}
\end{figure}

Figure~\ref{fig:projections} shows the result of the fit projected onto
the $m_{ES}$ and $\DeltaE$ observables. 
The histograms show the data after a cut on the quantity
${\cal P}_{\rm sig}/({\cal P}_{\rm sig}+{\cal P}_{\rm back})$ 
has been applied, where
${\cal P}_{\rm sig}$ and ${\cal P}_{\rm back}$ are the probabilities 
for a given event
to be signal and background respectively, and 
are evaluated using all the 
observables except the one that is being plotted. The cut is optimized 
for each variable separately. The solid (dashed) line shows the 
projection for the full fit (background only) after the same cut is applied.

\clearpage

\section{SYSTEMATIC STUDIES}
\label{sec:Systematics}

Table \ref{tab:syst} summarizes the systematic errors on the 
branching fraction. We include the uncertainty in the total number of 
\B mesons,
the uncertainties coming from selection cuts, such as track multiplicity,
thrust angle, and vertex requirements, and the uncertainty in particle
identification.
The uncertainty on the selection
efficiency is estimated from Monte Carlo and is dominated 
by the reconstruction of soft tracks. The accuracy of the simulation
of the key observables \mes and \DeltaE is estimated using
the  $\Bz\to\Dm\pip\to\Kp\pim\pim\pip$ control sample.

The peaking $\BB$ background has been taken into account in the fit with
three fixed components. The effect of their systematic
uncertainty on the signal is evaluated by adding to the data known samples of
background Monte Carlo events and observing the variation of the result.

Monte Carlo studies show that the interference between signal and
$a_1^{\pm}\pi^{\mp}$
may not be neglected and could significantly bias
the measurement. 
Assuming for the branching fraction of the two modes the central
values measured without accounting for interference, and  accounting for acceptance
and likelihood fit, the systematic bias is estimated at the level of
7.5 events.

To obtain the error associated to the uncertainty in the PDFs parameters, 
200 Monte Carlo experiments are performed with the parameters varied within 
their errors.
The width of the fitted yield distribution is taken as systematic error.
Finally, we assign a  systematic error of $\pm$ 3 events to cover a possible
fit bias evaluated with Monte Carlo experiments. 

\begin{table}[htb]
  \centering
  \caption{Summary of systematic errors in the measurement of the
\Bztorhozrhoz branching fraction.} 
\vspace{0.3cm}
  \begin{tabular}{lcc}  
\hline
Source                          &  \multicolumn{2}{c}{Uncertainty}  \\
	         		&  Yield fraction & Number of events \\
\hline                          
                                &  \multicolumn{2}{c}{Multiplicative}  \\   
\hline
Number of B mesons              &  1.1\%        &  0.4  \\
Track multiplicity cut          &  1\%          &  0.3  \\
Thrust angle cut                &  1\%          &  0.3  \\
Vertex requirement              &  2\%          &  0.7  \\
PID cut                         &  2\%          &  0.7  \\
Track finding                   &  3.2\%        &  1.1  \\
MC statistics                   &  $<$1\%       &  $<$0.3  \\
\hline                          
                                &  \multicolumn{2}{c}{Additive}  \\    
\hline
B background                    &  17.4\%       &  5.8  \\
$\mathrm{a}_1\pi$ interference  &  22.5\%       &  7.5   \\ 
PDF variation                   &  18\%         &  6.0  \\ 
Fit bias                        &  9\%          &  3.0   \\
\hline
                                &               & \vspace*{-0.4cm}  \\
Total                           &  35\%         & 11.7  \\
                                &               & \vspace*{-0.4cm}  \\
\hline
  \end{tabular}
  \label{tab:syst}
\end{table}

Taking systematic uncertainties into account, the measured value for
the $B\rightarrow \rho^0\rho^0$ branching fraction is

$${\cal B}(B\rightarrow \rho^0\rho^0) =
\left (0.54 ^ {+0.36}_{-0.32} 
\pm {0.19}
\right ) \times 10^{-6} $$

\noindent or an upper limit of $1.1 \times 10^{-6}$ at 90\% confidence level.

\section{SUMMARY}
\label{sec:Summary}
With a sample of 227 million \upsbb decays collected with 
the \babar\ detector we have searched for the decay mode 
\Btozz. We set an upper limit of $1.1\times 10^{-6}$
(90\% C.L.) on the branching fraction of this decay mode.
Our results are preliminary. This result has important 
implications for understanding the penguin contribution 
and constraints on the CKM angle $\alpha$ with 
the $B\to\rho\rho$ decays.
With the \Bztorhoprhom values of the branching fraction and 
longitudinal polarization~\cite{Aubert:2003xc},
the measured \Bztorhoprhom S and C \CP--violating time--dependent
asymmetry parameters~\cite{rhoprhom}, the 
\Bptorhozrrhop values of the branching fraction and 
longitudinal 
polarization~\cite{Aubert:2003mm}~\cite{Zhang:2003up}~\cite{HFAG}, 
and the measured branching ratio of \Bztorhozrhoz measured in this analysis, 
neglecting electroweak penguins, non-resonant and $I=1$ isospin 
contributions, and using isopin analysis~\cite{rhoprhom}, 
we obtain a new value for $\alpha$:

$$\alpha = \left ( 96 \pm 10 ({\rm stat}) 
\pm 4 ({\rm syst}) \pm 11 ({\rm penguin}) \right ) 
^{\mathrm o}$$

\begin{figure}[htb]
\begin{center}
\epsfig{figure=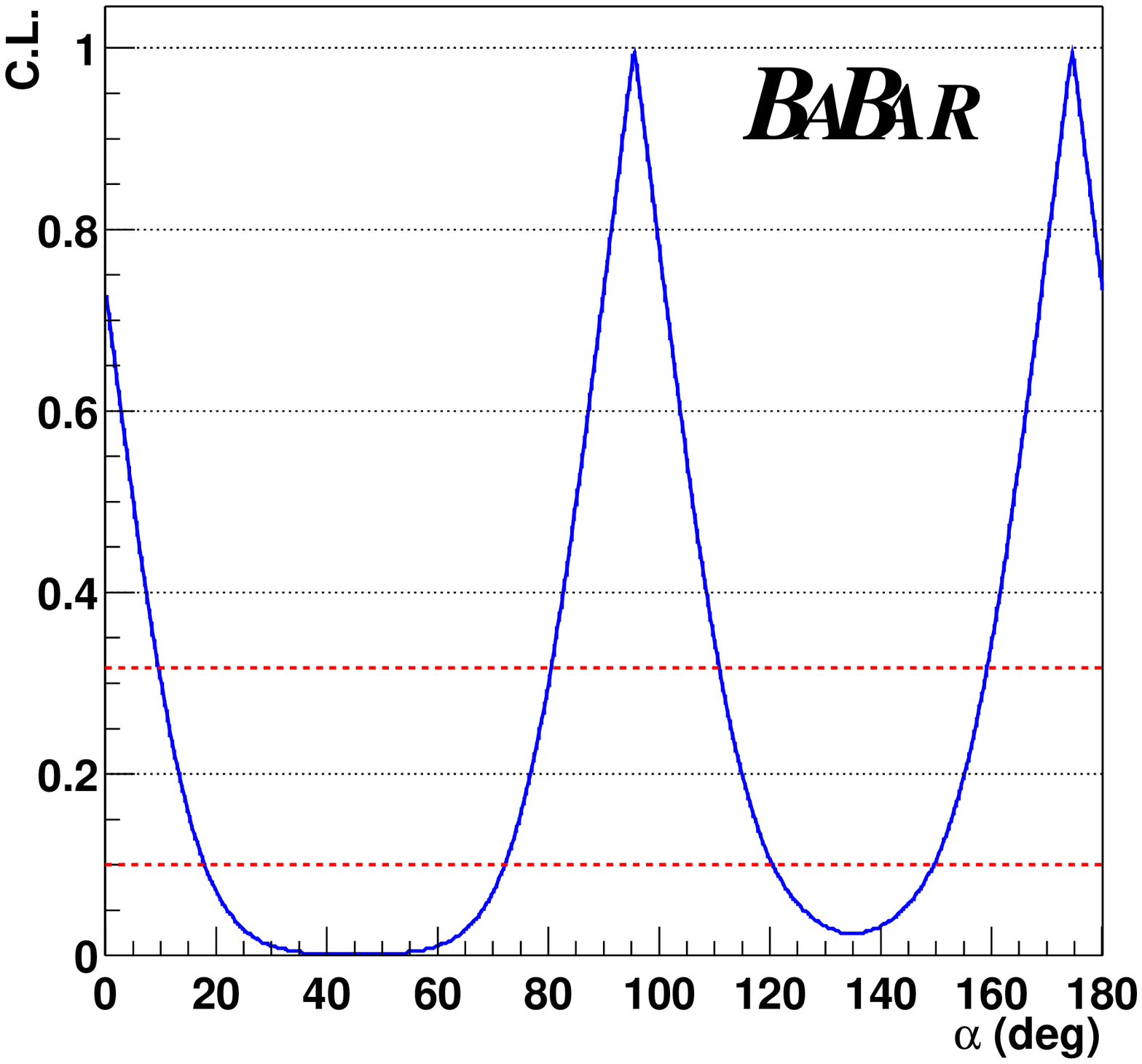,height=.4\textwidth}
\caption{Confidence Level on $\alpha$ obtained from the measured branching
fraction, fraction of longitudinal polarization, and \CP parameters of 
$\Bztorhoprhom$,
branching fraction and fraction of longitudinal polarization of
$\Bptorhozrrhop$, and
the $\Bztorhozrhoz$ branching fraction measurement
from this analysis, using isospin analysis. 
The dashed lines correspond to the
68\% (top) and 90\% (bottom) confidence intervals.}
\label{fig:alphascan}
\end{center}
\end{figure}

To extract alpha a $\chi^2$ in which all the measured quantities and the
angles $2\Delta\alpha^{+-}$ and $2\Delta\alpha^{00}$ of 
Fig.~\ref{fig:triangle} are expressed as a function of the length of 
the two isospin triangles' sides is mimimized. The confidence
level on $\alpha$ is obtained by a scan of the difference between 
$\chi^2(\alpha)$ for a given value of $\alpha$ and the minimum of 
this $\chi^2$. Fig.~\ref{fig:alphascan}
presents the result of this scan.

\vspace*{-8.9cm}
\hspace*{8.2cm} {\footnotesize {\sf Preliminary} \normalsize}
\vspace*{8.4cm}

\section{ACKNOWLEDGMENTS}
\label{sec:Acknowledgments}

\input pubboard/acknowledgements

\end{document}

%% file: pubboard/authors_sum2004.tex
\begin{center}
\small

The \babar\ Collaboration,
\bigskip

%
B.~Aubert,
R.~Barate,
D.~Boutigny,
F.~Couderc,
J.-M.~Gaillard,
A.~Hicheur,
Y.~Karyotakis,
J.~P.~Lees,
V.~Tisserand,
A.~Zghiche
\inst{Laboratoire de Physique des Particules, F-74941 Annecy-le-Vieux, France }
% ...

\end{center}\newpage

%% file: pubboard/acknowledgements.tex
We are grateful for the 
extraordinary contributions of our \pep2\ colleagues in
achieving the excellent luminosity and machine conditions
that have made this work possible.
The success of this project also relies critically on the 
expertise and dedication of the computing organizations that 
support \babar.
The collaborating institutions wish to thank 
SLAC for its support and the kind hospitality extended to them. 
This work is supported by the
US Department of Energy
and National Science Foundation, the
Natural Sciences and Engineering Research Council (Canada),
Institute of High Energy Physics (China), the
Commissariat \`a l'Energie Atomique and
Institut National de Physique Nucl\'eaire et de Physique des Particules
(France), the
Bundesministerium f\"ur Bildung und Forschung and
Deutsche Forschungsgemeinschaft
(Germany), the
Istituto Nazionale di Fisica Nucleare (Italy),
the Foundation for Fundamental Research on Matter (The Netherlands),
the Research Council of Norway, the
Ministry of Science and Technology of the Russian Federation, and the
Particle Physics and Astronomy Research Council (United Kingdom). 
Individuals have received support from 
CONACyT (Mexico),
the A. P. Sloan Foundation, 
the Research Corporation,
and the Alexander von Humboldt Foundation.